# Improving Performance of Organic Photodetectors based on Trap Assisted Photomultiplication through Adjusting Photogenerated Carrier Distributions

Wenyan Wang,[a] Linlin Shi,[a] Ye Zhang,[a] Yuying Hao,[a,b] Furong Zhu,[a,b] Kaiying Wang,[a] and Yanxia Cui[a,b,*]

[a]*Key Laboratory of Advanced Transducers and Intelligent Control System, Ministry of Education and Shanxi Province, College of Physics and Optoelectronics, Taiyuan University of Technology, Taiyuan 030024, China*

[b]*Key Laboratory of Interface Science and Engineering in Advanced Materials, Taiyuan University of Technology, Taiyuan 030024, China*

*Corresponding author: yanxiacui@gmail.com (Y. Cui)

ABSTRACT

Organic photodetectors (OPDs) based on trap-assisted photomultiplication (PM) effect with external quantum efficiency (EQE) far exceeding 100% are quite appealing for achieving highly sensitive photodetection. A classic structure of PM-type OPDs is based on the active layer of P3HT:PC$_{70}$BM with the weight ratio of 100:1, in which PC$_{70}$BM forms islands and supply bulk electron traps for inducing strong interfacial band bending and thus hole tunneling from the Al electrode. In this paper, aiming for optimizing the PM effect, we study the photogenerated carrier distribution by tuning thickness of the active layer (P3HT:PC$_{70}$BM). The combination effect of both the exciton generation and exciton dissociation processes affects the photogenerated carrier distribution, which ultimately determines the PM performances of OPDs. On the one hand, simulation reveals that the thinner the active layer, the stronger exciton generation near the Al electrode. On the other hand, the photoluminescence and surface morphology studies reflect that the active layer with thickness of 230 nm has the smooth surface and produces the highest exciton dissociation rate. The two effects result in the OPD with a 205 nm thick active layer has the champion EQE (105569%) and photoresponsivity (344 A/W), corresponding to an enhancement of 330% with respect to the OPD with a 325 nm thick active layer. Ascribed to the trade off of EQE and dark current, the detectivity performs differently at different wavelength ranges. At short wavelength range, the detectivity reaches the highest when the active layer is 205 nm thick, while the highest detectivity at long wavelength range is produced by the thickest active layer. Our work contributes to developing low cost organic photodetectors for detecting weak light signals.

## 1. Introduction

Recently, organic photodetectors (OPDs) [1-4] have received tremendous interests because of their unique advantages of abundant resource, light weight, good flexibility, etc. Particularly, OPD devices based on photomultiplication (PM) effect with external quantum efficiency (EQE) far exceeding unity can realize highly sensitive photodetection[5-12], which is significant for detecting weak light signal in many fields, such as bio-imaging sensing and long range light communication[13-15]. The working mechanism of PM type OPDs has been identified as the trap assisted carrier tunneling effects[8, 16-18]. For such effects, the optoelectronic conversion process includes four steps: light absorption in active layer, carrier generation/capture by traps after light illumination, carrier transport toward the Schottky junction under applied bias, and carrier tunneling through the Schottky barrier.

PM type OPDs were firstly invented using small organic molecules but their fabrication required expensive thin film deposition systems[4, 19-21]. In comparison, the wet solution processable polymers with heterojunctions inside are better candidates for developing low cost PM type OPDs[22-27]. Among the diverse heterojunction formulas, polymer and fullerene derivatives with weight ratio much higher than 1:1 are quite intensively investigated[10, 28-30]. In these kind of devices, the fullerene derivative acceptor forms isolated islands which can trap the photogenerated carriers and thereby induce the current multiplication. Its pioneer prototype with a configuration of ITO/PEDOT:PSS/P3HT:$PC_{71}BM$(100:1)/LiF/Al was demonstrated in 2015.[31], with an EQE of 16700% at -19 V bias. Later, the EQE was raised up to 37500% by lowering the hole injection barrier between the Al electrode and the active layer through removing the LiF buffer layer[8]. In the following, by avoiding atomic self-assembly, they obtained face-on arranged P3HT molecules which could facilitate the hole transportation, bringing forward the EQE rise higher to 115800% at -19 V[24].

In this work, we further optimize the performance of PM type OPDs through changing the thickness of the active layer. Our theoretical studies reflect that with thin active layer, more excitons are induced under light illumination near the Al electrode, thus an improved EQE of the trap assisted PM type OPDs might be expected. The experimental studies demonstrate that with the decrease of the thickness of the active layer from 325 nm to 205 nm thick, the EQE performance is gradually improved and further decrease of the active layer results in EQE degradation. The highest EQE of 105569% is realized at 405 nm wavelength under -17 V bias with a 205 nm thick active layer, corresponding to an enhancement of 330% compared with that of the OPD with a 325 nm thick active layer. The following photoluminescence (PL) and surface morphology studies of films indicate that the active layer with different thicknesses possess different exciton dissociation rates and the 230 nm thick film has the optimal exciton dissociation rate. The combination of both the photogenerated exciton distribution and the exciton dissociation rate ultimately determines the photogenerated carrier distribution in the active layer, explaining why the OPD with the champion EQE has a 205 nm thick active layer. What we also noticed, however, was that decreasing the thickness of the active layer also exacerbated the dark current ($J_d$). Overall, the PM type OPDs with a 205 nm thick active layer possess the maximum detectivity ($D^*$) of $1.87\times10^{13}$ Jones at wavelength range shorter than 500 nm and the device with a 325 nm thick active layer has the maximum $D^*$ of $2.32\times10^{13}$ Jones at wavelength range longer than 500 nm. Our work contributes to developing low cost and highly sensitive photodetectors.

## 2. Device structure and working mechanism

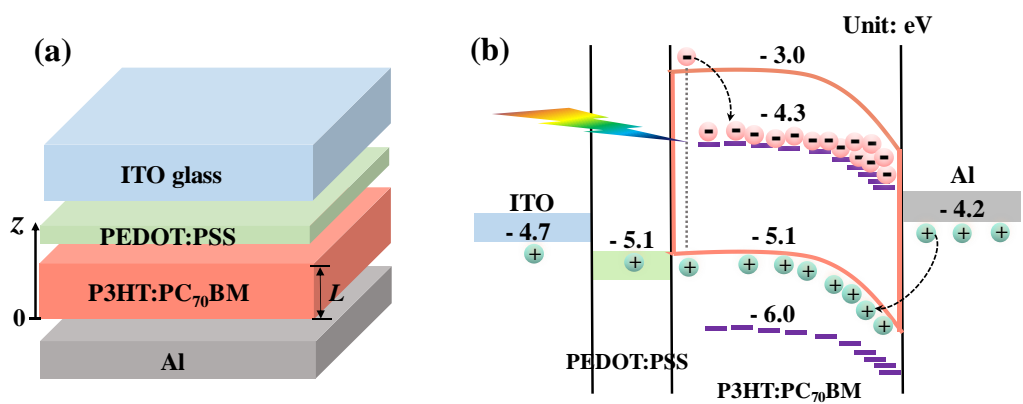

Fig. 1. (a) Device configuration of the reported PM type OPD and (b) its energy band diagram.

Fig. 1 shows structural and energy band sketch of the PM type OPD devices with configuration of ITO/PEDOT:PSS/P3HT:PC$_{70}$BM/Al. Both the ITO anode and Al cathode layers have thicknesses of 100 nm. The thickness of hole transport layer PEDOT:PSS is 25 nm. The active layer is made of P3HT:PC$_{70}$BM with the weight ratio of 100:1, in which the nano-sized islands of PC$_{70}$BM molecules are dispersed in the P3HT matrix. In our study, the thickness of the active layer ($L$) varies from 180 nm to 325 nm.

Fig. 1(b) displays the corresponding energy band diagram of the OPD device. The difference between the lowest unoccupied molecular orbital (LUMO) levels of P3HT (-3.0 eV) and PC$_{70}$BM (-4.3 eV) is 1.3 eV. With such a high energy barrier, the photogenerated electrons will be successfully trapped in the PC$_{70}$BM islands. Moreover, the hole injection barrier is about 0.9 eV between the highest occupied molecular orbital (HOMO) levels of Al and P3HT, therefore the holes are hardly injected from the Fermi level of Al (−4.2 eV) cathode into the HOMO level of P3HT (−5.1 eV). However, under illumination, holes might tunnel from the Al electrode to P3HT due to the existence of electron traps in the active layer. When a bias is applied with the electric field pointing from Al to ITO, the photogenerated electrons trapped by the PC$_{70}$BM islands will transport toward the P3HT:PC$_{70}$BM/Al junction, resulting in a large amount of electrons aggregate near the Al electrode. The aggregation of electrons will induce interfacial band bending and a corresponding narrower barrier, which finally results in stronger hole tunneling into the P3HT HOMO level, i.e., a higher current multiplication factor.

## 3. Simulation section

### 3.1 *Simulation methods*

The PM type OPD devices are investigated theoretically by the two-dimensional (2D) Finite Element Method (FEM) which had been verified in our previous study[32, 33]. All simulations have been carried out with periodic boundary conditions applied along the $z$ axis which is normal to the film plane. Perfectly matched layer (PML) boundaries are applied at two planes perpendicular to the $z$ axis, one of which lies in the glass and the other is in the air region next to the Al electrode. Light are illuminated from the ITO glass side. The wavelength dependent refractive indices ($n$) of P3HT:PC$_{70}$BM (100:1) are obtained from Ref. [29]. And other refractive indices of the materials used in this work are extracted from Refs. [34, 35]. The exciton generation originate from the absorption of photons, so we calculated the absorption efficiency of the whole active layer ($\eta$) versus wavelength ($\lambda$), based on which we obtained the integrated absorption efficiency ($\eta_I$) over the wavelength range between 300 nm and 650 nm.

3.2 *Simulation results and discussion*

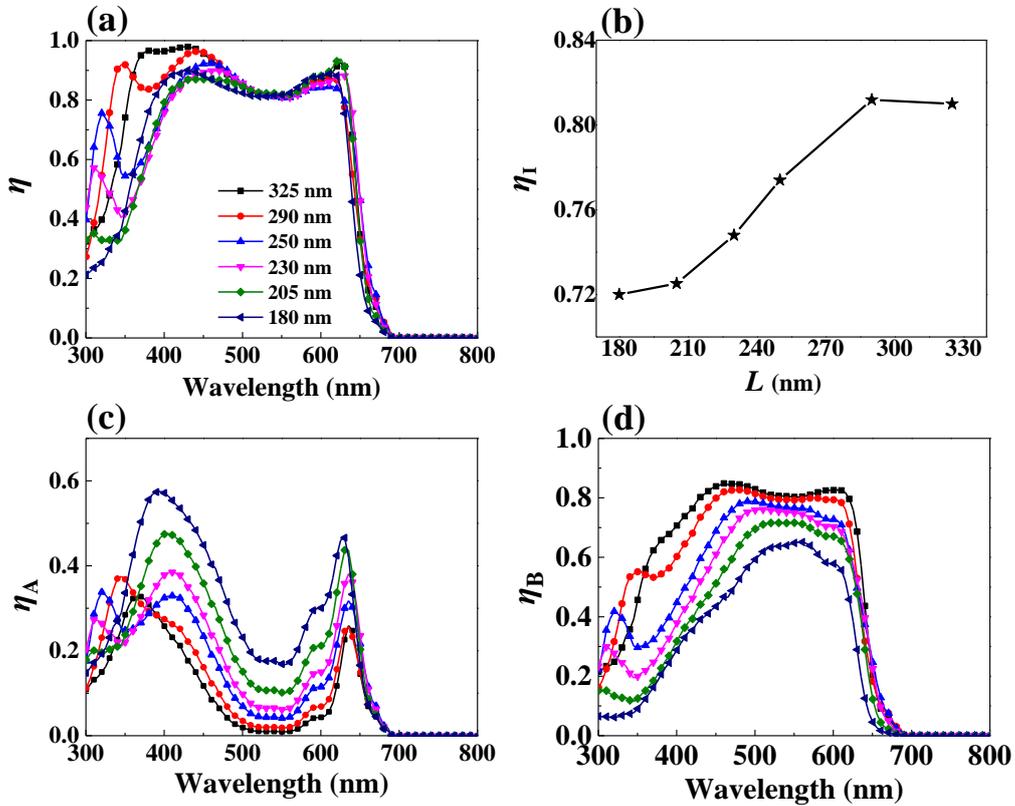

Fig. 2. (a) The absorption efficiency of the whole active layer ($\eta$) versus wavelength and (b) the corresponding integrated absorption efficiency ($\eta_I$) over the wavelength range between 300 nm and 650 nm when the thickness of the active layer ($L$) is tuned. (c-d) The wavelength dependent absorption efficiency in part A ($\eta_A$) and part B ($\eta_B$) of the active layer as $L$ varies. Part A is the region within 100 nm from the Al electrode and Part B is the remaining region as indicated in Fig. 3(f).

Figs. 2(a) and 2(b) show the curves of $\eta \sim \lambda$ at varied $L$ and the curve of $\eta_I \sim L$, respectively. It is seen clearly in Fig. 2(b) that when $L$ increases from 180 nm to 290 nm, $\eta_I$ increases gradually and further increase of $L$ leads to the reduction of $\eta_I$. Such tendency is because the absorption peak at the wavelength range shorter than 450 nm redshift with gradually enhanced intensity with the increase of $L$ as shown in Fig. 2(a). The local maxima of $\eta_I$ at $L = 290$ nm reflects a very large amount of the incident light is absorbed by the active material. But such thick active layer may not be favorable in the application of PM type OPDs because the photogenerated electrons in the region far away from the Al electrode could hardly arrive at the Al electrode due to poor electron transport property of the P3HT:PC$_{70}$BM(100:1) layer thereby contributing negligibly to strengthen the hole tunneling from the external circuit. Therefore, it is necessary to further examine the absorption maps.

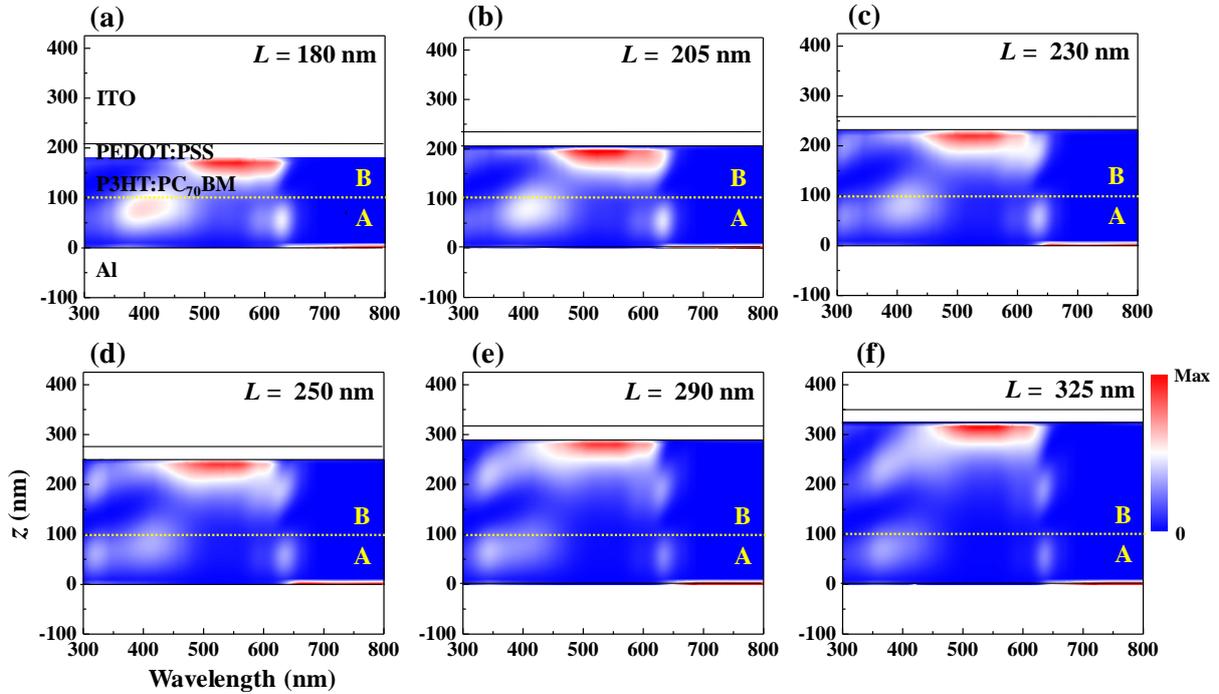

Fig. 3. Wavelength dependent maps of absorption inside of the active layer. (a) $L = 180$ nm, (b) $L = 205$ nm, (c) $L = 230$ nm, (d) $L = 250$ nm, (e) $L = 290$ nm, (f) $L = 325$ nm.

Fig. 3 shows the maps of the wavelength dependent absorption distributions of the active layer with thickness from 180 nm to 325 nm. Those maps exhibit several bright spots due to the Fabry-Pérot interference effect [36], which correspond to efficient absorption by the active material. The most bright interference spots centered at wavelength of ~ 525 nm are located near the PEDOT:PSS layer, and its distance from the Al electrode becomes smaller when the active layer is thinner. In addition, although the active layer with $L$ = 290 nm allows more interference bright spots spread out inside it at wavelengths shorter than 450 nm and longer than 600 nm, the intensities of its interference spots close to the Al electrode are much weaker, with respect to the thin active layer cases (e.g., $L$ = 180 nm or 205 nm). To quantify the absorption intensity near the Al electrode, we then divided the active layer into two parts. The region within 100 nm from the Al electrode is labelled as Part A and the remaining region is Part B as shown in Fig. 3. Then, the absorption of light in Part A ($\eta_A$) and Part B ($\eta_B$) versus wavelength are calculated, respectively, at varied $L$, as displayed in Figs. 2(c) and 2(d). Obviously, it can be seen, although the absorption in the whole active layer is weakened in thin active layers, the absorption in Part A increases gradually with the decrease of the thickness of the active layer. Instead, the absorption in Part B decreases gradually when the active layer becomes thicker. From the simulation results, we conclude that the thickness of the active layer determines the absorption distributions and therefore the exciton generation distributions, playing a great influence on the performances of OPDs.

## 4. Experiment section

### 4.1 *Experiment details*

The indium tin oxide (ITO) coated glass substrates with a sheet resistance of 15 Ω/square were sequentially cleaned by ultrasonic treatment in detergent, deionized water, and ethanol for 15 min, respectively. Then, all cleaned ITO substrates were dried by air and treated in the surface plasma cleaner for 5 min to increase the work function of ITO substrates. The solution of PEDOT:PSS (purchased from Xi'an p-OLED) was spin-coated onto the cleaned ITO substrates at a spin speed of 5000 rounds per minute (rpm) for 30 s, which was baked in air at 120 °C for 15 min. Polymer P3HT and PC$_{70}$BM (purchased from Xi'an p-OLED and 1-Material Corporation, respectively) were dissolved in 1,2-dichlorobenzene (o-DCB) to prepare 40 mg/ml mixed solutions at the weight ratio of 100:1. The mixed solutions were spin-coated onto the PEDOT:PSS films at a certain spin speed for 30 s in high nitrogen-filled glove box, then the samples are annealed for 20 s at 80 °C. The spin speed of 800 rpm, 1000 rpm, 1200 rpm, 1400 rpm, 1600 rpm, and 1800 rpm produced the active layer with thicknesses of 325 nm, 290 nm, 250 nm, 230 nm, 205 nm, and 180 nm, respectively. The relationship between the spin speed and the thickness of the active layer is almost linear. Then, the 100 nm thick aluminum electrode was thermally evaporated on the active layer in a high vacuum (10-4 Pa) chamber. The active area of each fabricated OPD device was about 0.04 cm$^2$, defined by the overlap of Al cathode and ITO anode.

The thicknesses of the spin-coated and evaporated films were characterized by profiler (Bruker, DektakXT). The current density–voltage (*J–V*) characteristics of OPDs in the dark and under illumination (with a burn-in treatment [9]) were measured using a programmable source meter (Keithley 2400). The monochromatic light source was supplied by a xenon lamp (ZOLIX GLORIA-X150A) combined with a monochromator (ZOLIX 0mni-$\lambda$ 3005) with the light intensities at varied wavelengths characterized by a power meter (Ophir NOVA II). The surface morphology of the P3HT:PC$_{70}$BM was characterized by the atomic force microscope (AFM) (Flying Man, Nanoview 1000). The absorption spectra of multilayer films were recorded using a Hitachi U-3900. The steady state photoluminescence (PL) spectra of films were recorded using a fluorescence spectrophotometer (Edinburgh Instruments, FLS-980).

4.2 *Definition of performance parameters*

The performance parameters of the fabricated PM type OPDs including external quantum efficiency, photoresponsivity, detectivity, and response time are defined as follows.

The external quantum efficiency is defined as the electron number detected per incident photon, as indicated by the following equation:

$$\text{EQE} = \frac{N_e}{N_p} = \frac{I_{ph}/e}{P_{in}/h\nu} \tag{1}$$

, where $N_e$ and $N_p$ are the number of detected electrons and incident photons, respectively, $I_{ph}$ is the photocurrent, $P_{in}$ is the incident light intensity, $h$ is the Planck's constant, $\nu$ is the frequency of light, and $e$ is the electronic charge.

In the investigated PM type OPDs, the presence of deep electron traps by the $PC_{70}BM$ islands causes a long electron recombination lifetime, resulting in a high photocurrent amplification gain, similar to that happens in photoconductor type photodetectors. In photodetectors with gain mechanisms, EQE is equal to the gain in number. Based on the definition of gain, EQE can also be given by Equation (2):

$$\text{EQE} = \frac{\chi\tau}{T} \tag{2}$$

, where $\chi$ is the fraction of trapped electrons over the total amount of the dissociated excitons, $\tau$ is the lifetime of trapped electrons, and $T$ is the transport time of the hole flowing across the active layers.

The photoresponsivity ($R$) is defined as the ratio of photocurrent to the power intensity of incident light, which can be expressed by Equation (3):

$$R = \frac{I_{ph}}{P_{in}} = \frac{I_l - I_d}{P_{in}} \tag{3}$$

, in which $I_l$ is the current under light illumination, and $I_d$ is the dark current.

The detectivity ($D^*$) is the figure of merit for quantifying the capability of weak light detection, which can be derived from the photoresponsivity and the noise density. Considering that the noise current under dark is dominated by shot noise, the detectivity can be calculated through Equation (4):

$$D^* = \frac{R}{\sqrt{2eJ_d}} \quad (4)$$

, in which $J_d$ is the dark current density.

4.3 *Experimental results and discussions*

The electrical measurements for different OPDs indicate that the thickness of the active layer indeed has significant influences on the performances of the trap assisted PM effects. Figs. 4(a) and 4(b) show the EQE and $R$ spectra, respectively, of different OPDs measured at −17 V bias. It is seen clearly that the EQE (and $R$) values of the OPDs with active layer thickness of 325 nm and 290 nm are the smallest two groups over all curves although their absorption abilities are quite strong as indicated by Fig. 2(a) as well as by the measured absorption spectra of the whole device (not shown). With the decrease of the active layer, the EQE and $R$ are reduced; and the device with $L = 205$ nm possess maximum EQE of 105569% and champion $R$ of 344 A/W obtained at wavelength of 405 nm, corresponding to an enhancement of around 330% with respect to those when $L = 325$ nm (EQE = 24875%, $R = 80$ A/W). An obvious demerit of the thinning of the active layer is the worsening of the dark current ($J_d$) as shown in Fig. 4(c), which is due to the lowering of the series resistance with the decrease of the thickness of the active layer. Ascribed to the trade off of EQE and $J_d$, the detectivity $D^*$ spectra of the OPDs have different performances at different wavelength ranges, as shown in Fig. 4(d). It is found that at wavelength range shorter than 500 nm, $D^*$ reaches the highest of $1.87 \times 10^{13}$ Jones with $L = 205$ nm, while at wavelength range longer than 500 nm, the highest $D^*$ is $2.32 \times 10^{13}$ Jones with $L = 325$ nm.

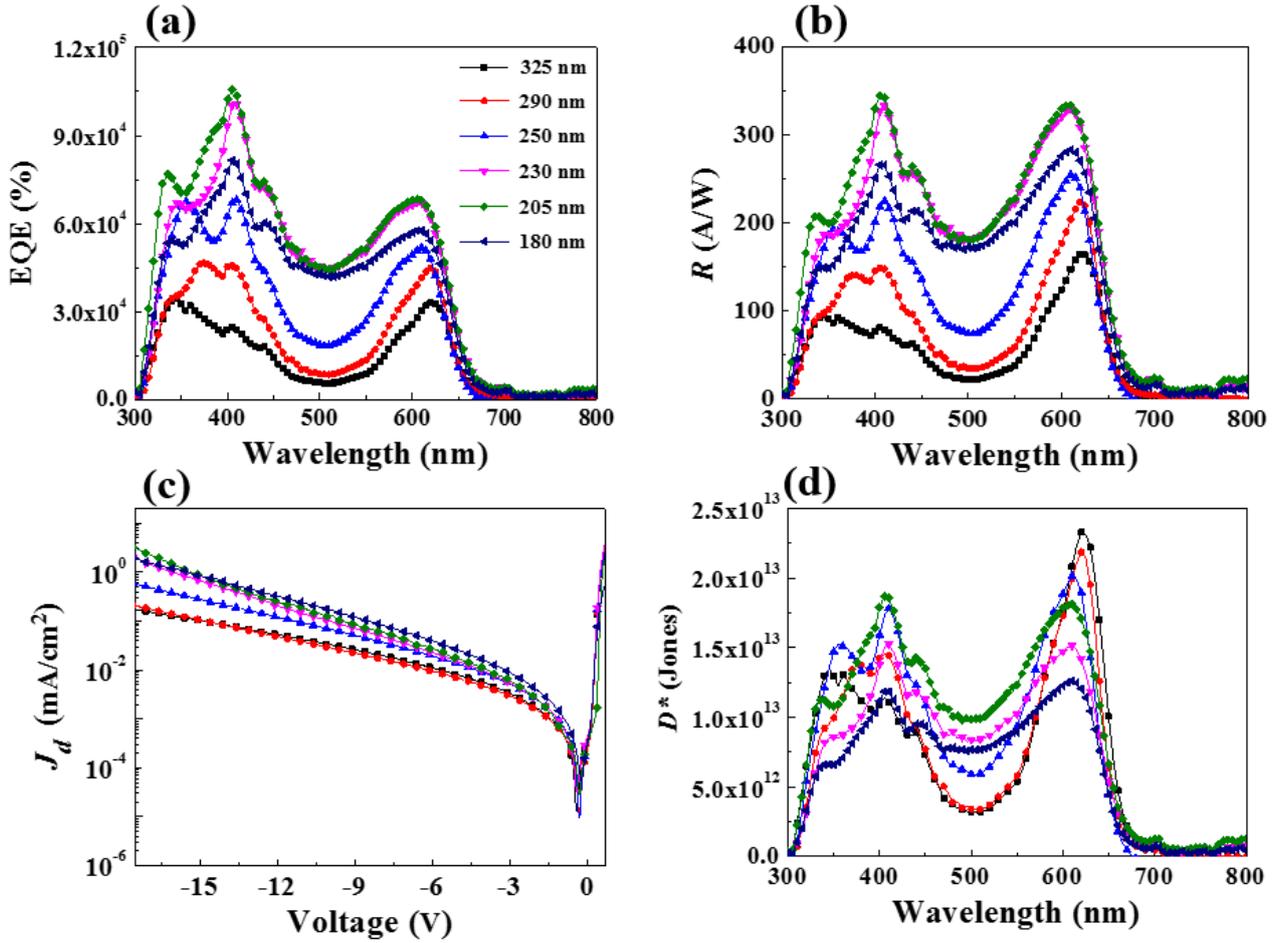

Fig. 4. Characteristics of OPD devices when $L$ is tuned. (a) The external quantum efficiency (EQE) spectra, (b) the photoresponsivity ($R$) spectra, (c) the dark current density ($J_d$) versus the bias, and (d) the detectivity ($D^*$) spectra.

The performances of EQE are explained according to the equation (2) of EQE. Here, on the assumption that $\tau$ is constant for different OPDs, both $\chi$ and $L$ can affect the EQE of OPDs. On one hand, the decrease of $L$ directly brings forward the decrease of $T$, causing an enhancement on EQE. On the other hand, the decrease of $L$ from 325 nm to 180 nm also increase the absorption of light near the Al electrode as indicated in Fig. 2(c), which should lead to a greater $\chi$ and thus an enhancement on EQE as well. However, as we observed, the EQE and photoresponsivity are optimized at $L = 205$ nm instead of the smallest $L$. To explain the observation, we further fabricated the multilayer films of ITO/PEDOT:PSS/P3HT:PC$_{70}$BM with varied active layer thicknesses and measured their absorption spectra and PL spectra, as displayed in Figs. 5(a) and 5(b), respectively. One sees from Fig. 5(a) that the absorption efficiencies over the broad wavelength range from 300 nm to 650 nm increase gradually with the increase of the thickness of the active layer from 180 nm to 290 nm. And the two absorption curves of $L = 290$ nm and 325 nm are more or less the same. The study of absorption spectra indicates that the P3HT:PC$_{70}$BM layer with a larger $L$ has a higher exciton generation rate, therefore a higher PL intensity would be expected if the exciton generation rates are the same for different films. However, it is interesting to find out from Fig. 5(b) that the PL intensity at the emission peak of ~730 nm is optimized when $L = 230$ nm, reflecting that the film has a higher exciton dissociation rate which limits the radiative transition of excitons. We then measured the surface morphologies of the P3HT:PC$_{70}$BM films and the results are shown in Fig .6. It is seen that, with the decrease of $L$ from 325 nm to 180 nm, the surface roughness (RMS) of the active layer first decreases and then increases. The corresponding RMS in Figs. 6(a)-6(f) are 3.47 nm, 2.59 nm, 2.10 nm, 1.44 nm, 2.11 nm, and 2.66 nm, respectively. The smallest RMS is obtained when $L = 230$ nm, of which the PL intensity is also the highest as shown in Fig. 5(b). This reflects that the enhanced exciton dissociation rate at $L = 230$ nm is due to the smoother film morphology. In combination with the exciton generation and exciton dissociation properties, the amount of photogenerated tapped electrons near the Al electrode (which relates to $\chi$ in the equation of EQE) get optimized at *L* greater than 180 nm. Therefore, although the decrease of *L* induces a smaller *T* and thus a greater EQE, the champion EQE values were achieved at *L* = 205 nm.

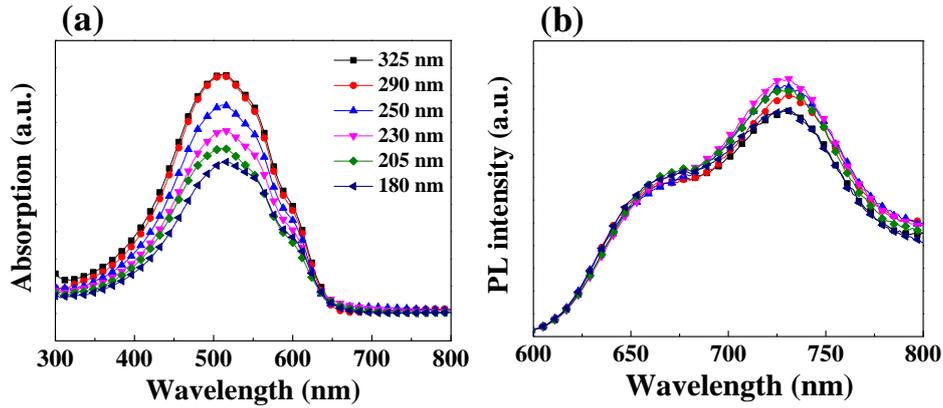

Fig. 5. (a) Absorption and (b) PL spectra of the multilayer films of ITO/PEDOT:PSS/ P3HT:PC$_{70}$BM when $L$ is tuned.

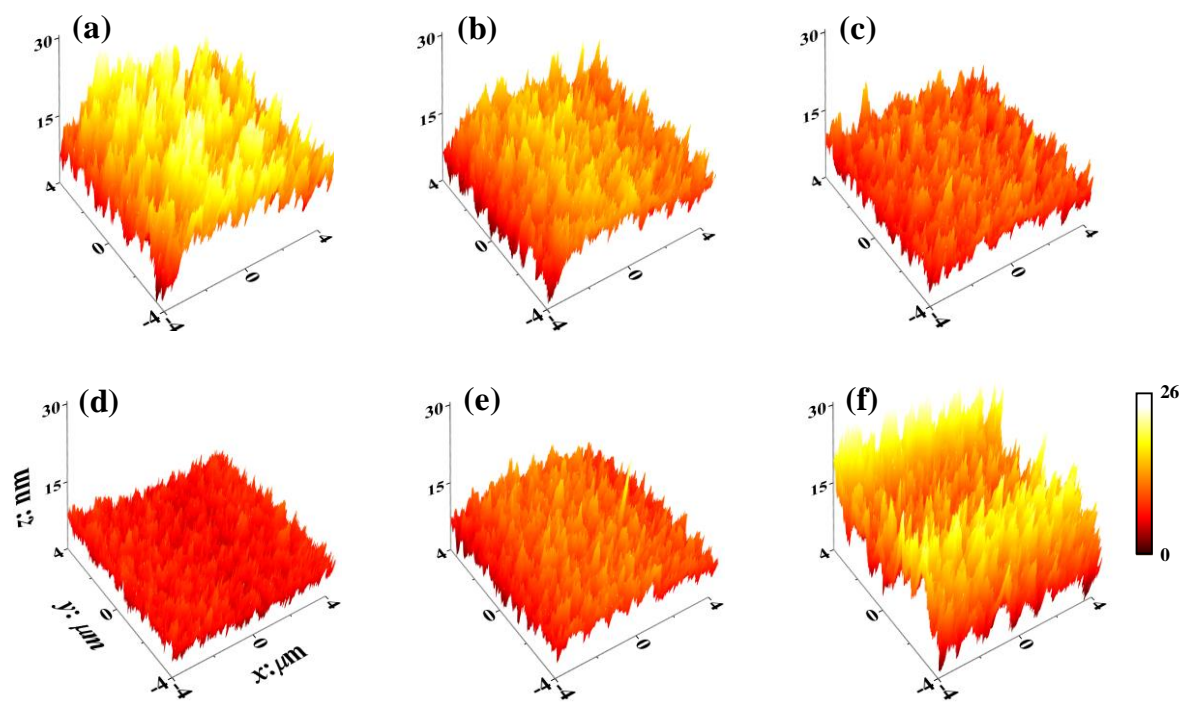

Fig. 6. AFM images of the ITO/PEDOT:PSS/ P3HT:PC$_{70}$BM films when *L* is tuned. (a) *L* = 325 nm, (b) *L* = 290 nm, (c) *L* = 250 nm, (d) *L* = 230 nm, (e) *L* = 205 nm, and (f) *L* = 180 nm.

## 5. Conclusions

PM-type OPDs based on the active layer P3HT:PC$_{70}$BM with a weight ratio of 100:1 have been systematically studied by tuning the active layer thickness. In theoretical calculation, the change of the active layer thickness shows vital influences on the distribution of light absorption, i.e., the distribution of exciton generation in the active layer. It is revealed that the thinner the active layer, the stronger exciton generation near the Al electrode. In the following experiment studies, it has been confirmed that the EQE spectral and the dark current of OPDs strongly depend on the thickness of the active layer. It is demonstrated that the highest EQE (105569%) was achieved at 405 nm wavelength under -17 V bias when the thickness of active layer is 205 nm. Through the absorption, photoluminescence, and surface morphology studies of the ITO/PEDOT:PSS/P3HT:PC$_{70}$BM(100:1) multilayer films, we have found that the active layer thickness also affects the exciton dissociation rate and the film with a 230 nm thick active layer possesses the highest exciton dissociation rate due to improved morphologies. Therefore, although the thinnest active layer has the strongest exciton generation near the Al electrode according to the simulation results, the amount of photogenerated electrons at that region is not optimal. It is the amount of the photogenerated electrons and the hole transport time through the active layer, both of which are influenced by the active layer thickness, codetermines the gain intensity, resulting in the champion EQE achieved when a 205 nm thick active layer is applied. In addition, because the lowest dark current was realized when the active layer of 325 nm thick is used, the corresponding detectivity reaches the highest of $1.87 \times 10^{13}$ Jones at wavelength range longer than 500 nm. But at wavelength range shorter than 500 nm, $D^*$ reaches the highest of $1.87 \times 10^{13}$ Jones with $L = 205$ nm. It is believed that the findings in this work can also be applied in other PM type OPDs. Our work is helpful for the development of low cost high performance organic photodetectors.

**Acknowledgements**

We are grateful for support from the National Natural Science Foundation of China (61775156, 61805172, 61571317, and U1710115), the Natural Science Foundation of Shanxi Province (201701D211002), Henry Fok Education Foundation Young Teachers fund, Young Sanjin Scholars Program, Key Research and Development (International Cooperation) Program of Shanxi Province (201603D421042), and Platform and Base Special Project of Shanxi Province (201605D131038). We also thank Fujun Zhang, Wenbin Wang, and Jianli Miao at Beijing Jiaotong University for their guidance on the part of experimental details.